\documentclass[letter,longauth]{aa}
\usepackage[utf8]{inputenc}
\usepackage{graphicx}
\usepackage{float}
\usepackage[colorlinks=true, 
            linkcolor=blue,  
            citecolor=blue,  
            filecolor=blue  
           ]{hyperref}

\usepackage{txfonts}
\usepackage{orcidlink}
\usepackage{siunitx}

\begin{document} 

   \title{Caught in the web: galaxy mergers along cosmic filaments}

   \author{Carolina Dulcien\orcidlink{0009-0003-4392-2283}
          \inst{1,2,3}
          \and
          Yara~L. Jaff\'e\orcidlink{0000-0003-2150-1130}\inst{2,3}\fnmsep\thanks{\email{yara.jaffe@usm.cl}}
          \and
          Jacob P. Crossett\orcidlink{0000-0002-9810-1664}
          \inst{2}
          \and
          Raúl Baier-Soto\orcidlink{0009-0008-4255-1309}
         \inst{2,3,4}
         \and
         Hugo~M\'endez-Hern\'andez\orcidlink{0000-0003-3057-9677}\inst{5,3}
         \and
        Christopher~P.~Haines\orcidlink{0000-0002-8814-8960}\inst{6,3}
        \and
        Guillermo~Cabrera-Vives\orcidlink{0000-0002-2720-7218}\inst{1,3}
        \and
        Patricio Olivares\orcidlink{0009-0003-4757-8627}\inst{2}
        \and
         P. Vásquez-Bustos\orcidlink{0000-0001-7140-3980}\inst{7}
         \and 
         Maria Argudo-Fernández\orcidlink{0000-0002-0789-2326}\inst{7,8}  
         \and
        Javiera Vivanco\orcidlink{0009-0009-1525-0313}\inst{2}
        \and
        Lawrence Bilton\orcidlink{0000-0002-4780-129X}\inst{9,10,11}
        \and
        Clécio R. Bom\orcidlink{0000-0003-4383-2969}\inst{11}
        \and
        Giuseppe D'Ago\orcidlink{0000-0001-9697-7331}\inst{12}
        \and
        Alexis Finoguenov\orcidlink{0000-0002-4606-5403}\inst{13}
        \and
        Ulrike Kuchner\orcidlink{0000-0002-0035-5202}\inst{14}
        \and
        Ciria~Lima-Dias\orcidlink{0009-0006-0373-8168}\inst{5}
        \and
        Paola~Merluzzi\orcidlink{0000-0003-3966-2397}\inst{13}
        \and
        Antonela~Monachesi\orcidlink{0000-0003-2325-9616}\inst{5}
        \and
        Diego~Pallero\orcidlink{0000-0002-1577-7475}\inst{2,3}
        \and
        Nicolas~Tejos\orcidlink{0000-0002-1883-4252}\inst{4}
        \and
        Gabriel S. M. Teixeira\orcidlink{0000-0002-1594-208X}\inst{11}
        \and
        Cristóbal Sifón\orcidlink{0000-0002-8149-1352} \inst{4}
        \and
        Maiara S. Carvalho\orcidlink{0009-0008-9042-4478} \inst{16}
        \and
        Ricardo~Demarco\orcidlink{0000-0003-3921-2177}\inst{17}
        \and
        Eduardo~Ibar\orcidlink{0009-0008-9801-2224} \inst{9,3}
        \and
        Gissel P. Montaguth\orcidlink{0009-0003-1364-3590}\inst{16}
        \and
        Franco Piraino-Cerda\orcidlink{0009-0008-0197-3337}\inst{2,3,4}
        \and
        Umberto Rescigno\orcidlink{0000-0002-9280-0173 }\inst{6}
        \and
        Vitor Sampaio\orcidlink{0000-0001-6556-637X}\inst{2,3}
        \and
        Gustavo B. Oliveira Schwarz\orcidlink{0009-0003-6609-1582}\inst{18}
        \and
        Rory Smith\orcidlink{0000-0001-5303-6830}\inst{2,3}
        Benedetta~Vulcani\orcidlink{0000-0003-0980-1499} \inst{13}
        \and
        Nicola~Malavasi\orcidlink{0000-0001-9033-7958}\inst{19}
          }

     \institute{
   Universidad de Concepción, Víctor Lamas 1290, Barrio Universitario, Concepción, Chile
   \and
   Departamento de Física, Universidad Técnica Federico Santa María, Avenida España 1680, Valparaíso, Chile
   \and
   Millennium Nucleus for Galaxies (MINGAL)
   \and 
   Instituto de Física, Pontificia Universidad Católica de Valparaíso, Casilla 4059, Valparaíso, Chile
   \and
   Departamento de Astronom\'ia, Universidad de La Serena, Avda. Raúl Bitr\'an 1305, La Serena, Chile
   \and
   Instituto de Astronomía y Ciencias Planetarias (INCT), Universidad de Atacama, Copayapu 485, Copiapó, Chile
   \and
   Departamento de Física Teórica y del Cosmos, Edificio Mecenas, Campus Fuentenueva, Universidad de Granada, E-18071 Granada, Spain
   \and 
   Instituto Universitario Carlos I de Física Teórica y Computacional, Universidad de Granada, 18071 Granada, Spain
   \and  
   Instituto de Física y Astronomía, Universidad de Valparaíso, Avda. Gran Bretaña 1111, Valparaíso, Chile
   \and 
   Centre of Excellence for Data Science, Artificial Intelligence \& Modelling, The University of Hull, Cottingham Road, Kingston-Upon-Hull, HU6 7RX, UK
   \and
   Centro Brasileiro de Pesquisas Físicas, Rua Dr. Xavier Sigaud 150, Rio de Janeiro, RJ, Brazil
   \and
   Institute of Astronomy, University of Cambridge, Madingley Road, Cambridge CB3 0HA, United Kingdom
   \and
   INAF–Osservatorio Astronomico di Capodimonte, Salita Moiariello 16, I-80134 Napoli, Italy
   \and
   Department of Physics, University of Helsinki, Gustaf Hällströmin katu 2, 00560 Helsinki, Finland
   \and
   School of Physics \& Astronomy, University of Nottingham, Nottingham NG7 2RD, UK
   \and
   Departamento de Astronomia, Instituto de Astronomia, Geofísica e Ciências Atmosféricas, Universidade de São Paulo, São Paulo, Brazil
   \and
   Institute of Astrophysics, Facultad de Ciencias Exactas, Universidad Andrés Bello, Sede Concepción, Talcahuano, Chile
   \and
   Escola Politécnica, Universidade de São Paulo, Av. Prof. Luciano Gualberto, Travessa do Politécnico 380, São Paulo, Brazil
   \and
    Max-Planck-Institut für extraterrestrische Physik (MPE), Gießenbachstraße 1, D-85748 Garching bei München, Germany
    }

   \date{Received March XX, 2025; accepted MXXX}

  \abstract
   {Galaxy clusters grow through the accretion of galaxies from groups, filaments, and other clusters. During this process, galaxies may undergo pre-processing in lower-density environments, where galaxy–galaxy mergers and other interactions can significantly alter their properties prior to cluster infall.
   }
   {We investigate the role of galaxy mergers in the pre-processing of galaxies prior to cluster infall by studying the spatial distribution of galaxy mergers across the cosmic web.}
   {We use a sample of 43,922 galaxies being targeted by the 4MOST CHANCES survey in and around 33 low-redshift clusters ($z<0.07$). Using Zoobot, a deep-learning framework trained on Galaxy Zoo data, we identify 698 galaxy mergers. We measure their distances to cosmic web filaments and compare it to those of non-merging galaxies. }
   {We find that galaxy mergers are significantly closer to filaments than the non-merging galaxy population, with this trend being strongest beyond the cluster virial radius. This suggests that filaments provide conditions conducive to mergers, possibly moderating relative velocities and enhancing gas availability. }
   {Our findings support a scenario in which filaments play a key role in transforming (pre-processing) galaxies by promoting mergers 
   before entering the cluster cores where star formation quenches. 
   }


   \maketitle
%

\section{Introduction}

Understanding galaxy evolution requires considering both intrinsic properties and the environment in which galaxies reside. In the current cosmological framework—the hierarchical structure formation paradigm—cosmic structures grow via the gravitational merging of smaller units over time. This process builds the large-scale structure (LSS) of the Universe: from galaxies to groups, and ultimately, massive galaxy clusters interconnected by filaments and separated by voids \citep{White1978, Press1974}. 
Observations provide compelling evidence for  hierarchical growth, where galaxies are accreted along filaments into larger systems \citep{Zwicky19333,Eke1996}.

A wide array of physical mechanisms influence galaxy properties throughout this hierarchical growth. These include internal processes, such as star formation and AGN feedback, as well as environmental effects like gas stripping via ram-pressure from the intracluster medium, tidal interactions, and galaxy-galaxy mergers \citep{boselli2006, Alonso2007, Darg2010, Weigel2018, Ellison2019}. Each of these can significantly alter a galaxy's morphology, star formation activity, and gas content, contributing to the broad diversity observed in the galaxy population today. 
Mergers can alter the kinematics of galaxies (including spin orientation), reshape galactic disks, trigger intense starburst events, and feed central supermassive black holes, potentially igniting AGN activity \citep{Toomre1972, Barnes1992, Hopkins2006, Cox2006, Welker2014, Barsanti2025}. 

These environmental mechanisms are expected to act in different locations of the cosmic web. In dense cluster cores for instance, galaxies experience significant gas removal through ram-pressure stripping \citep{Jaffe2015, Cortese2021}. However, a substantial fraction of galaxies enter clusters already quenched, suggesting that they undergo “pre-processing” in lower-density environments prior to cluster infall \citep[][]{Fujita2004,Haines2015,Lopes2023}. 

In intermediate-density environments (such as galaxy groups or the outskirts of clusters), mechanisms such as galaxy-galaxy mergers are expected to be more effective, as the lower relative velocities facilitate gravitational binding \citep{boselli2006}.  Recent observational results using deep learning classifications \citep{omori2023} show that merger incidence increases in lower-density environments on scales of \(0.5\text{--}0.8~h^{-1}\,\mathrm{Mpc}\)  compared to dense environments. This is supported by simulations, which indicate that galaxy mergers preferentially occur in environments with lower relative velocities, such as intermediate-mass haloes, rather than in massive cluster cores \citep{Jian2012}.

Despite this framework, observational studies of mergers at filament-cluster interfaces remain limited, with the specific role of mergers in the pre-processing of galaxies along filaments unexplored systematically.

In this work, we explore the role of galaxy mergers in the pre-processing of galaxies falling into clusters. As the main channel of galaxy accretion into clusters is through large-scale filaments \citep{Cautun2014}, we focus our analysis on the distribution of galaxy mergers relative to filaments feeding clusters. Specifically, we investigate whether galaxy mergers tend to occur along filamentary structures and how this tendency varies with distance from the cluster centre.
Throughout this paper, $R_{200}$ refers to the radius within which the mean density is 200 times the critical density at the cluster’s redshift. We adopt a Planck 2020 cosmology with $H_0 = 67.4~\mathrm{km\,s^{-1}\,Mpc^{-1}}$ and $\Omega_m = 0.315$.

\vspace{-5pt}
\section{Dataset}

\subsection{Cluster and Galaxy sample}

This paper focuses on galaxies in clusters at $z<0.07$ from the Low-z sub-survey of the 4MOST CHileAN Cluster galaxy Evolution Survey \citep[CHANCES; ][]{Haines2023, sifón2024chanceschileanclustergalaxy}. This survey is tailored to study galaxy pre-processing and will obtain $\sim$500,000 spectra of galaxies within and around more than 100 clusters out to $5 \times R_{200}$ at $0<z<0.45$ with the 4MOST instrument on the 4-m VISTA telescope \citep{dejong2019}, down to low stellar masses ($m_r < 20.4$~mag).

\subsection{Large scale environment of galaxies}
To characterize galaxy environment we focus on galaxies out to $5 \times R_{200}$ from the cluster centre, selected using photometric redshifts  from the DESI Legacy Surveys Data Release 10 (LS--DR10; \citealt{Zhou2021}), following the CHANCES target selection strategy, which selects cluster galaxies in redshift slices centered on each cluster redshift \citep[see][for details]{mendez2026}. For the environmental characterization, we limit the sample to galaxies brighter than  $m_r = 18.5$~mag\footnote{This limit corresponds to $\log_{10}(M_*/M_{\odot}) \sim 9$}, where the completeness and purity in the selection of cluster members is high ($\sim 80 \%$) thanks to the reduced photometric redshift uncertainty
\citep[$\langle \sigma_z \rangle \simeq 0.145$ for LS--DR10, and 
$\langle \sigma_z \rangle \simeq 0.044$ for our improved LS--DR10--CBPF estimates; see][]{mendez2026}. Using this magnitude range, \citet{baier2025} has shown already that filaments can be adequately identified with photometric members.

Filament identification was performed in 2D using the Discrete Persistence Structures Extractor \cite[DisPerSE\footnote{\url{https://www2.iap.fr/users/sousbie/disperse.html}}; ][]{Sousbie2011},
which is widely used for detecting cosmic web components in observations \citep{Bonjean2020} and simulations \citep{Galárraga-Espinosa2020,Kuchner2020,Kraljic2020, Malavasi2022}. 
We applied boundary conditions for smooth Delaunay tessellation, with persistence threshold at $3\sigma$ and smoothness level of 20 following \cite{baier2025}. 
Galaxy–filament distances ($D_{fil}$) were computed via the point-line method (See Appendix~\ref{app:A}) along with the projected distance to the cluster centre ($r_{cl}$).

\subsection{Galaxy morphology}  

To obtain morphological information for the photometric members, we cross-matched the CHANCES catalogue of target galaxies with catalogues containing data from Galaxy Zoo DESI \citep[GZDESI; ][]{walmsley2023galaxyzoodesidetailed}, which provides automated morphology measurements based on deep images from 
the DESI Legacy Imaging Surveys DR8 \citep[DESI-LS-DR8; ][]{dey2019}. The GZDESI morphologies are derived using Zoobot \citep{zoobot}, a deep learning model trained to replicate the consensus of the Galaxy Zoo volunteer classifications. For each galaxy, Zoobot predicts vote fractions: the expected proportion of human classifiers who would select a given morphological label (e.g., "smooth" or "featured").

\begin{figure*}
    \centering
    \includegraphics[width=1\textwidth]{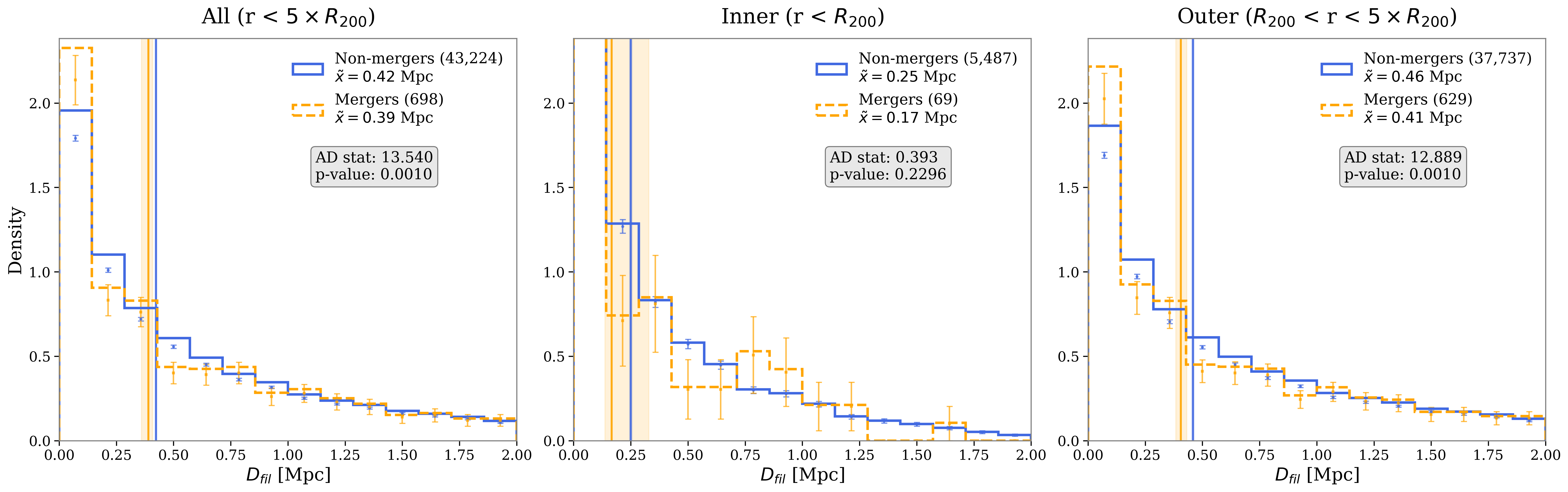}
    \caption{Normalized distribution of $D_{fil}$ for all merging galaxies (orange dashed line), compared to the non-merging galaxy population (blue dotted line). Left: Distribution for galaxies across the entire cluster region (r < $5\times R_{200}$);  Middle: galaxies in the  inner cluster region (r < $R_{200}$). Right: galaxies in the cluster outskirts ($R_{200}$ < r < $5\times R_{200}$).  For visualization, distances to filaments are truncated at \(2\,\mathrm{Mpc}\), beyond which the galaxy density drops consistently below \(\sim 0.2\). Error bars correspond to Poisson uncertainties in each bin, while vertical lines mark the median of each distribution. Shaded regions indicate the 68\% bootstrap confidence intervals (percentiles 16–84) around the median. The AD statistics and p-values comparing both distributions are shown in the legend. The AD statistics show that galaxy mergers are preferentially located closer to filaments than non-merging galaxies, both at all radii and specifically outside the virial region. Distances to filaments are truncated at 2~Mpc in the plots for visibility, although the full distributions extend up to $\sim$4~Mpc.
    } 
    \label{fig:all}
\end{figure*}

While not all galaxies in the CHANCES Low-z survey had a match in the morphological catalogues of GZDESI (since CHANCES is based on DESI LS-DR10 rather than DR8 used for GZDESI), 43,922 galaxy matches were found spread across 33 different clusters (note that 8 clusters are not fully covered by GZDESI, see blue points in Fig.~\ref{fig:all_mosaic}). 
  
This sample of galaxies represents the general galaxy population (which includes both merging and non-merging galaxies).

To identify galaxies undergoing mergers using the GZDESI classifications we consider the vote fractions predicted by Zoobot in the following categories: merger (M-M), major disturbance (M-D), minor disturbance, and none. We visually inspected galaxies with different vote fractions for a random sub-sample of 2,000 galaxies to define a threshold that allowed us to build a reliable galaxy-galaxy merger sample. We adopted the following thresholds:  

 $F_{\rm MM} \geq 0.5 \hspace{0.1cm}$ or $\hspace{0.1cm} F_{\rm MD} \geq 0.5$, 

where $F_{\rm MM}$ and $F_{\rm MD}$ correspond to the predicted vote fractions for the M-M and M-D cases respectively\footnote{Note that our conclusions are insensitive to the exact merger probability threshold, remaining stable for values between $F=0.3$–$0.6$. Higher thresholds are primarily limited by small-number statistics.}. These thresholds yield a high purity (78\%) at the expense of completeness (45\%), which translates into a selection of clear cases of major mergers or disturbances. To ensure this level of purity, we excluded galaxies with minor disturbances or no signs of interaction, which would dilute the sample.

A total of 698 galaxy mergers were identified, corresponding to $\sim$2\% of the sample, implying that the remaining 43,224 galaxies are classified as non-mergers.
This merger fraction is low, but expected in cluster environments \citep[see e.g.][]{Kim_2024}, and even more so given our conservative criteria for selecting mergers.
Figure~\ref{fig:mosaic} shows a few randomly selected example galaxies classified as mergers.

\section{Results: Distribution of Mergers across the cosmic web}

To determine whether galaxy mergers contribute to the pre-processing of galaxies in filaments before they enter clusters, we compare their spatial distribution in the cosmic web to that of the general galaxy population (43,922 galaxies). This reference sample includes mergers, allowing us to test whether they exhibit distinct environmental preferences using the Anderson-Darling k-sample (AD) test.

Figure \ref{fig:all} shows the distribution of $D_{fil}$ for all merging galaxies compared to the general galaxy population, across different regions: the entire region analyzed ($r < 5 \times R_{200}$), in the cluster region ($r < R_{200}$), and in the cluster outskirts ($R_{200} < r <5 \times R_{200}$). 
In two cases, galaxy mergers are preferentially located closer to the filaments than the general galaxy population, with the difference being more pronounced outside the virialized region of the cluster, where merging galaxies have a median filament distance of $0.406^{+0.029}_{-0.024}$\,Mpc, compared to $0.459^{+0.005}_{-0.004}$\,Mpc for the non-merging galaxies, as shown in the right panel of Figure~\ref{fig:all}. The quoted uncertainties correspond to the 16th--84th percentiles derived from 1000 bootstrap resamplings ($1\sigma$ confidence intervals).

While the difference in the medians is small, the difference in the shape of the distributions has statistical significance according to the AD test which yields p-values well below 0.07\footnote{We employ a significance threshold of $\alpha = 0.07$ for our AD tests to account for the inherent observational bias toward filamentary environments present in both samples. This threshold enhances our sensitivity to detect merger-specific spatial deviations while maintaining statistical rigor (false-positive rate of 7\%). 
} outside the virial region. 
Considering that the typical filament thickness is $\sim 0.7 \ -\  1$ Mpc \citep[e.g.,][]{tempel2014detecting,Kuchner2020} this result suggests that mergers are more likely to occur within the filamentary structures themselves. The fact that the trend is more pronounced in the cluster outskirts is not surprising, as the relative velocities between galaxies there are low enough to allow mergers. 
Hence in the rest of the analysis, we consider only the outer regions of clusters. However, we assessed possible edge effects on the filament detections by re-doing the analysis considering an area limited to r<$4.8\times R_{200}$, and we obtained the same results and significance.

We also note that the mass distributions of the merger and non-merger samples are very similar but not statistically identical. Nonetheless, we have checked that our results do not change when matching the stellar mass distributions (see Appendix \ref{app:C}).

When examining the distribution of mergers cluster-by-cluster (shown in Figure~\ref{fig:outside}), we found statistically suggestive evidence (p < 0.07) that mergers preferentially occur near filaments in $\sim$39\% of our sample in the outer regions. For the remaining clusters, although visual inspection suggests similar distribution patterns to those with statistically significant results, the limited number of identified mergers prevents these trends from reaching statistical significance. 

\begin{figure}
    \centering
    \includegraphics[width=0.35\textwidth]{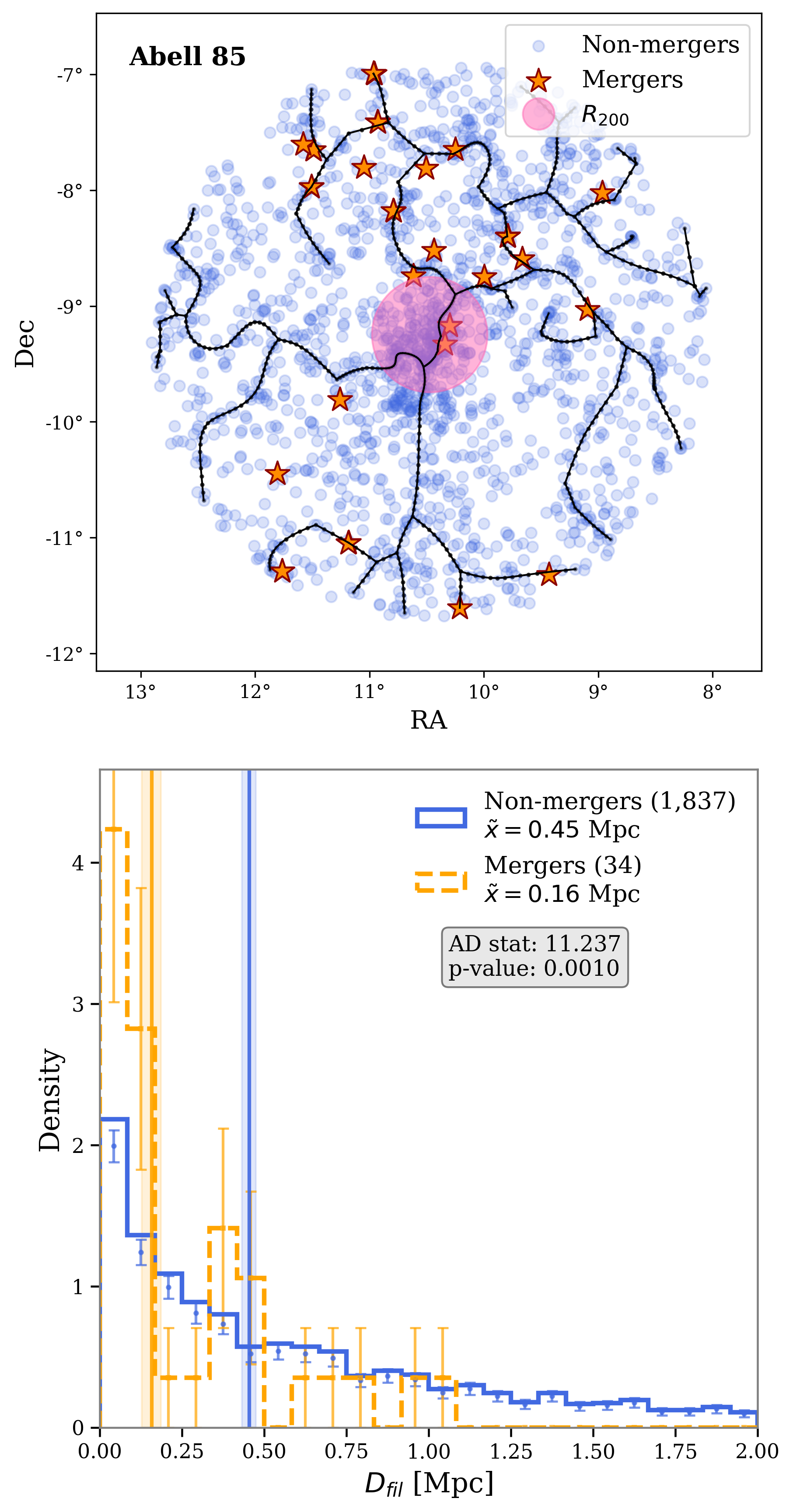}
    \caption{Distribution of galaxy mergers in the example cluster Abell 85. Top: Spatial distribution of galaxies in the cluster out to $5 R_{200}$ showing merging galaxies (orange stars) and non-merging galaxies (blue dots) relative to intracluster filaments (black lines). The pink circle denotes the $R_{200}$ radius for reference. Bottom: Normalized distribution of $D_{fil}$, comparing merging galaxies (orange) with the non-merging galaxy population (blue) at r < $5\times R_{200}$. Symbols are as in Figure~\ref{fig:all}. In this cluster mergers are significantly closer to filaments by a factor of 2.6 compared with non-merging galaxies. 
    } 
    \label{fig:85}
\end{figure}

Abell 85, a massive relaxed cluster in our sample, provides the strongest example of merger-filament connection, as shown in Figure \ref{fig:85}. In this cluster, mergers clearly trace the filamentary structure, supported by statistical indicators including a p-value of 0.001 and a median distance difference of 0.29 Mpc between merging galaxies and the general population.

Our results suggest that filamentary structures play an important role in galaxy evolution by facilitating merger events that may contribute to the pre-processing of galaxies before they fully enter the cluster environment.

\vspace{-5pt}
\section{Discussion}
\vspace{-0.1em}

We find a clear spatial association between galaxy mergers and the filamentary structure of the cosmic web—the first direct observational evidence of enhanced merger activity along cluster-feeding filaments. This connection is strongest beyond the virial radius, where lower relative velocities and the filamentary geometry likely promote gravitational interactions. Thus, filaments not only channel galaxies into clusters but also serve as active sites of pre-processing through mergers.

Our findings agree with simulations predicting enhanced merger activity, morphological transformation, and spin reorientation within filaments \citep{Welker2014, Dubois2014, Kuutma2017, Singh2020}, and they confirm previous indirect observational hints \citep{Malavasi2016, Mesa_2018, omori2023, rong2024}.

Our analysis is nevertheless subject to several limitations. Filament reconstruction relies on projected galaxy distributions and photometric redshifts, so line-of-sight uncertainties and shot noise may blur filament positions and dilute the signal. Moreover, our conservative merger selection prioritizes purity over completeness, likely underestimating the true merger incidence. Future studies using spectroscopy and more detailed merger identification will help confirm our results.

\section{Conclusions}

We have investigated the role of cosmic filaments in promoting galaxy mergers as part of the pre-processing of galaxies prior to their infall into clusters. Using a sample of ${\sim}$\,40,000 galaxies in and around 33 low-redshift ($z < 0.07$) clusters from the CHANCES survey, we study the spatial distribution of mergers across the cosmic web. Our main finding is that galaxy mergers are preferentially located near filaments outside the virial region of clusters.

These results provide evidence that cosmic filaments are not merely passive conduits of galaxy infall but active sites of transformation, where mergers contribute to the pre-processing of galaxies prior to cluster infall. 

In future work, we aim to expand this analysis using the full CHANCES sample through a dedicated citizen science project that incorporates additional physical mechanisms at play. We will also enrich the analysis with 4MOST spectroscopy,  more complex local and global environmental metrics, and comparisons with cosmological hydrodynamical simulations, which can help constrain the underlying physical drivers of filament-driven evolution. 
Moreover, we will explore whether mergers preferentially occur within galaxy groups/substructures embedded in filaments, comparing the impact of local and LSS-environment, once CHANCES provides robust group/substructures catalogues and central/satellite classifications.

\begin{acknowledgements}
      YLJ and RS acknowledge support from Agencia Nacional de Investigaci\'on y Desarrollo (ANID) through FONDECYT Regular projects 1241426 and 1230441. YLJ, CD, HMH, CPH, RBS, FPC, EI, VMS, RS and DP gratefully acknowledge financial support from the Millennium Science Initiative Program NCN2024\_112. YLJ, RBS, HMH, AM, RD gratefully acknowledge support from the ANID BASAL project FB210003. RBS acknowledges support from ANID/Subdirección de Capital Humano/Doctorado Nacional/2023-21231017.  
      PVB and MAF acknowledge support from the research project PID2023-150178NB-I00 and PID2023-149578NB-I00, financed by MCIN/AEI/10.13039/501100011033, the project A-FQM-510-UGR20 financed from FEDER/Junta de Andalucía-Consejería de Transformación Económica, Industria, Conocimiento y Universidades/Proyecto, by the grants P20\_00334 and FQM108, financed by the Junta de Andalucía (Spain), the Emergia program (EMERGIA20\_38888) from Junta de Andalucía, and the Grant AST22-4.4, funded by Consejería de Universidad, Investigación e Innovación and Gobierno de España and Unión Europea - NextGenerationEU. CPH acknowledges support from ANID through Fondecyt Regular project number 1252233. HMH acknowledges support from ANID through Fondecyt project 3230176. AM acknowledges support from the ANID FONDECYT Regular grant 1251882, and funding from the HORIZON-MSCA-2021-SE-01 Research and Innovation Programme under the Marie Sklodowska-Curie grant agreement number 101086388. MSC acknowledges support from São Paulo Research Foundation (FAPESP) grant 2023/10774-5 and funding to the S-PLUS project through the process 2019/26492-3. EI gratefully acknowledges financial support from ANID FONDECYT Regular 1221846. CLD acknowledges support from the ANID through Fondecyt project 3250511. GD acknowledges support by UKRI-STFC grants: ST/T003081/1 and ST/X001857/1. We thank A.~J. Koch-Hansen, S.~Barsanti and A.~Fritz for useful comments.

\end{acknowledgements}

\bibliographystyle{aa}
\bibliography{refs.bib}

@ARTICLE{mendez2026,
       author = {{M{\'e}ndez-Hern{\'a}ndez}, Hugo and {Lima-Dias}, Ciria and {Monachesi}, Antonela and {Jaff{\'e}}, Yara L. and {Haines}, Christopher P. and {Teixeira}, Gabriel S.~M. and {L{\"o}sch}, Elismar and {Baier-Soto}, Ra{\'u}l and {Lima}, Erik V.~R. and {Amrutha}, B.~M. and {Bom}, C.~R. and {D'Ago}, Giuseppe and {Demarco}, Ricardo and {Finoguenov}, Alexis and {Haack}, Rodrigo F. and {Lopes}, Amanda R. and {de Oliveira}, C. Mendes and {Merluzzi}, Paola and {Piraino-Cerda}, Franco and {Smith Castelli}, Anal{\'\i}a V. and {Sif{\'o}n}, Cristobal and {Sodr{\'e}}, Jr., Laerte and {Tejos}, Nicol{\'a}s and {Torres-Flores}, Sergio and {Argudo-Fern{\'a}ndez}, Maria and {Crossett}, Jacob P. and {Ibar}, E. and {Kuchner}, Ulrike and {Lacerna}, Ivan and {Lopes-Silva}, Vitor H. and {Lopez}, Sebasti{\'a}n and {McGee}, Sean and {Morelli}, Lorenzo and {Nantais}, Julie and {Olivares V.}, Patricio and {Pallero}, Diego and {Poggianti}, Bianca M. and {Pompei}, Emanuela and {Sampaio}, V.~M. and {Vulcani}, Benedetta and {Zenteno}, Alfredo and {Almeida-Fernandes}, F. and {Bilicki}, Maciej and {Carvalho}, M.~S. and {Cheng}, Cheng and {Figueiredo},, A.~L. and {Guti{\'e}rrez-Soto}, L.~A. and {Herpich}, F.~R. and {Kanaan}, A. and {Lacerda}, E.~A.~D. and {Nakazono}, L. and {Oliveira Schwarz}, G.~B. and {Ribeiro}, T. and {Roukema}, Boudewijn F. and {Sartori}, Mar{\'\i}lia J. and {Santos-Silva}, Tha{\'\i}s and {Schoenell}, W.},
        title = "{Targeting cluster galaxies for the 4MOST CHANCES Low-z sub-survey with photometric redshifts}",
      journal = {\aap},
         year = 2026,
        month = jan,
       volume = {706},
          eid = {A34},
        pages = {A34},
          doi = {10.1051/0004-6361/202556796},
archivePrefix = {arXiv},
       eprint = {2510.19958},
 primaryClass = {astro-ph.GA},
       adsurl = {https://ui.adsabs.harvard.edu/abs/2026A&A...706A..34M}
}

@ARTICLE{deJong2019,
       author = {{de Jong}, R.~S. and {Agertz}, O. and {Berbel}, A.~A. and {Aird}, J. and {Alexander}, D.~A. and {Amarsi}, A. and {Anders}, F. and {Andrae}, R. and {Ansarinejad}, B. and {Ansorge}, W. and {Antilogus}, P. and {Anwand-Heerwart}, H. and {Arentsen}, A. and {Arnadottir}, A. and {Asplund}, M. and {Auger}, M. and {Azais}, N. and {Baade}, D. and {Baker}, G. and {Baker}, S. and {Balbinot}, E. and {Baldry}, I.~K. and {Banerji}, M. and {Barden}, S. and {Barklem}, P. and {Barth{\'e}l{\'e}my-Mazot}, E. and {Battistini}, C. and {Bauer}, S. and {Bell}, C.~P.~M. and {Bellido-Tirado}, O. and {Bellstedt}, S. and {Belokurov}, V. and {Bensby}, T. and {Bergemann}, M. and {Bestenlehner}, J.~M. and {Bielby}, R. and {Bilicki}, M. and {Blake}, C. and {Bland-Hawthorn}, J. and {Boeche}, C. and {Boland}, W. and {Boller}, T. and {Bongard}, S. and {Bongiorno}, A. and {Bonifacio}, P. and {Boudon}, D. and {Brooks}, D. and {Brown}, M.~J.~I. and {Brown}, R. and {Br{\"u}ggen}, M. and {Brynnel}, J. and {Brzeski}, J. and {Buchert}, T. and {Buschkamp}, P. and {Caffau}, E. and {Caillier}, P. and {Carrick}, J. and {Casagrande}, L. and {Case}, S. and {Casey}, A. and {Cesarini}, I. and {Cescutti}, G. and {Chapuis}, D. and {Chiappini}, C. and {Childress}, M. and {Christlieb}, N. and {Church}, R. and {Cioni}, M. -R.~L. and {Cluver}, M. and {Colless}, M. and {Collett}, T. and {Comparat}, J. and {Cooper}, A. and {Couch}, W. and {Courbin}, F. and {Croom}, S. and {Croton}, D. and {Daguis{\'e}}, E. and {Dalton}, G. and {Davies}, L.~J.~M. and {Davis}, T. and {de Laverny}, P. and {Deason}, A. and {Dionies}, F. and {Disseau}, K. and {Doel}, P. and {D{\"o}scher}, D. and {Driver}, S.~P. and {Dwelly}, T. and {Eckert}, D. and {Edge}, A. and {Edvardsson}, B. and {Youssoufi}, D.~E. and {Elhaddad}, A. and {Enke}, H. and {Erfanianfar}, G. and {Farrell}, T. and {Fechner}, T. and {Feiz}, C. and {Feltzing}, S. and {Ferreras}, I. and {Feuerstein}, D. and {Feuillet}, D. and {Finoguenov}, A. and {Ford}, D. and {Fotopoulou}, S. and {Fouesneau}, M. and {Frenk}, C. and {Frey}, S. and {Gaessler}, W. and {Geier}, S. and {Gentile Fusillo}, N. and {Gerhard}, O. and {Giannantonio}, T. and {Giannone}, D. and {Gibson}, B. and {Gillingham}, P. and {Gonz{\'a}lez-Fern{\'a}ndez}, C. and {Gonzalez-Solares}, E. and {Gottloeber}, S. and {Gould}, A. and {Grebel}, E.~K. and {Gueguen}, A. and {Guiglion}, G. and {Haehnelt}, M. and {Hahn}, T. and {Hansen}, C.~J. and {Hartman}, H. and {Hauptner}, K. and {Hawkins}, K. and {Haynes}, D. and {Haynes}, R. and {Heiter}, U. and {Helmi}, A. and {Aguayo}, C.~H. and {Hewett}, P. and {Hinton}, S. and {Hobbs}, D. and {Hoenig}, S. and {Hofman}, D. and {Hook}, I. and {Hopgood}, J. and {Hopkins}, A. and {Hourihane}, A. and {Howes}, L. and {Howlett}, C. and {Huet}, T. and {Irwin}, M. and {Iwert}, O. and {Jablonka}, P. and {Jahn}, T. and {Jahnke}, K. and {Jarno}, A. and {Jin}, S. and {Jofre}, P. and {Johl}, D. and {Jones}, D. and {J{\"o}nsson}, H. and {Jordan}, C. and {Karovicova}, I. and {Khalatyan}, A. and {Kelz}, A. and {Kennicutt}, R. and {King}, D. and {Kitaura}, F. and {Klar}, J. and {Klauser}, U. and {Kneib}, J. -P. and {Koch}, A. and {Koposov}, S. and {Kordopatis}, G. and {Korn}, A. and {Kosmalski}, J. and {Kotak}, R. and {Kovalev}, M. and {Kreckel}, K. and {Kripak}, Y. and {Krumpe}, M. and {Kuijken}, K. and {Kunder}, A. and {Kushniruk}, I. and {Lam}, M.~I. and {Lamer}, G. and {Laurent}, F. and {Lawrence}, J. and {Lehmitz}, M. and {Lemasle}, B. and {Lewis}, J. and {Li}, B. and {Lidman}, C. and {Lind}, K. and {Liske}, J. and {Lizon}, J. -L. and {Loveday}, J. and {Ludwig}, H. -G. and {McDermid}, R.~M. and {Maguire}, K. and {Mainieri}, V. and {Mali}, S. and {Mandel}, H.},
        title = "{4MOST: Project overview and information for the First Call for Proposals}",
      journal = {Msngr},
         year = 2019,
        month = mar,
       volume = {175},
        pages = {3-11},
          doi = {10.18727/0722-6691/5117},
archivePrefix = {arXiv},
       eprint = {1903.02464},
 primaryClass = {astro-ph.IM},
       adsurl = {https://ui.adsabs.harvard.edu/abs/2019Msngr.175....3D}
}

@article{Malavasi2016,title={The VIMOS Public Extragalactic Redshift Survey (VIPERS): galaxy segregation inside filaments at z ≃ 0.7},author={N. Malavasi and S. Arnouts and D. Vibert and S. D. Torre and T. Moutard and C. Pichon and I. Davidzon and K. Kraljic and M. Bolzonella and L. Guzzo and L. Guzzo and B. Garilli and M. Scodeggio and B. Granett and B. Granett and U. Abbas and C. Adami and D. Bottini and A. Cappi and A. Cappi and O. Cucciati and O. Cucciati and P. Franzetti and A. Fritz and A. Iovino and J. Krywult and V. Brun and O. Fèvre and D. Maccagni and K. Małek and F. Marulli and F. Marulli and M. Polletta and M. Polletta and M. Polletta and A. Pollo and L. Tasca and R. Tojeiro and D. Vergani and A. Zanichelli and J. Bel and E. Branchini and J. Coupon and G. Lucia and Y. Dubois and A. Hawken and O. Ilbert and C. Laigle and L. Moscardini and L. Moscardini and T. Sousbie and M. Treyer and G. Zamorani},journal={\mnras},year={2016},volume={465},pages={3817-3822},doi={10.1093/mnras/stw2864}}

@ARTICLE{Kuutma2017,
       author = {{Kuutma}, Teet and {Tamm}, Antti and {Tempel}, Elmo},
        title = "{From voids to filaments: environmental transformations of galaxies in the SDSS}",
      journal = {\aap},
         year = 2017,
        month = apr,
       volume = {600},
          eid = {L6},
        pages = {L6},
          doi = {10.1051/0004-6361/201730526},
archivePrefix = {arXiv},
       eprint = {1703.04338},
 primaryClass = {astro-ph.GA},
       adsurl = {https://ui.adsabs.harvard.edu/abs/2017A&A...600L...6K}
}

@article{Dubois2014,title={Dancing in the dark: galactic properties trace spin swings along the cosmic web},author={Y. Dubois and C. Pichon and C. Welker and D. L. Borgne and J. Devriendt and C. Laigle and S. Codis and D. Pogosyan and S. Arnouts and K. Benabed and E. Bertin and J. Blaizot and F. Bouchet and J. Cardoso and S. Colombi and V. Lapparent and V. Desjacques and R. Gavazzi and S. Kassin and T. Kimm and H. McCracken and B. Milliard and S. Peirani and S. Prunet and S. Rouberol and J. Silk and A. Slyz and T. Sousbie and R. Teyssier and L. Tresse and M. Treyer and D. Vibert and M. Volonteri},journal={\mnras},year={2014},volume={444},pages={1453-1468},doi={10.1093/mnras/stu1227}}

@article{Singh2020,title={Study of galaxies on large-scale filaments in simulations},author={Ankit Singh and S. Mahajan and J. Bagla},journal={\mnras},year={2020},volume={497},pages={2265-2275},doi={10.1093/mnras/staa1913}}

@ARTICLE{Taylor2011,
       author = {{Taylor}, Edward N. and {Hopkins}, Andrew M. and {Baldry}, Ivan K. and {Brown}, Michael J.~I. and {Driver}, Simon P. and {Kelvin}, Lee S. and {Hill}, David T. and {Robotham}, Aaron S.~G. and {Bland-Hawthorn}, Joss and {Jones}, D.~H. and {Sharp}, R.~G. and {Thomas}, Daniel and {Liske}, Jochen and {Loveday}, Jon and {Norberg}, Peder and {Peacock}, J.~A. and {Bamford}, Steven P. and {Brough}, Sarah and {Colless}, Matthew and {Cameron}, Ewan and {Conselice}, Christopher J. and {Croom}, Scott M. and {Frenk}, C.~S. and {Gunawardhana}, Madusha and {Kuijken}, Konrad and {Nichol}, R.~C. and {Parkinson}, H.~R. and {Phillipps}, S. and {Pimbblet}, K.~A. and {Popescu}, C.~C. and {Prescott}, Matthew and {Sutherland}, W.~J. and {Tuffs}, R.~J. and {van Kampen}, Eelco and {Wijesinghe}, D.},
        title = "{Galaxy And Mass Assembly (GAMA): stellar mass estimates}",
      journal = {\mnras},
         year = 2011,
        month = dec,
       volume = {418},
       number = {3},
        pages = {1587-1620},
          doi = {10.1111/j.1365-2966.2011.19536.x},
archivePrefix = {arXiv},
       eprint = {1108.0635},
 primaryClass = {astro-ph.CO},
       adsurl = {https://ui.adsabs.harvard.edu/abs/2011MNRAS.418.1587T}
}

@ARTICLE{Kraljic2020,
       author = {{Kraljic}, Katarina and {Pichon}, Christophe and {Codis}, Sandrine and {Laigle}, Clotilde and {Dav{\'e}}, Romeel and {Dubois}, Yohan and {Hwang}, Ho Seong and {Pogosyan}, Dmitri and {Arnouts}, St{\'e}phane and {Devriendt}, Julien and {Musso}, Marcello and {Peirani}, S{\'e}bastien and {Slyz}, Adrianne and {Treyer}, Marie},
        title = "{The impact of the connectivity of the cosmic web on the physical properties of galaxies at its nodes}",
      journal = {\mnras},
         year = 2020,
        month = jan,
       volume = {491},
       number = {3},
        pages = {4294-4309},
          doi = {10.1093/mnras/stz3319},
archivePrefix = {arXiv},
       eprint = {1910.08066},
 primaryClass = {astro-ph.GA},
       adsurl = {https://ui.adsabs.harvard.edu/abs/2020MNRAS.491.4294K}
}

@ARTICLE{Malavasi2022,
       author = {{Malavasi}, Nicola and {Langer}, Mathieu and {Aghanim}, Nabila and {Gal{\'a}rraga-Espinosa}, Daniela and {Gouin}, C{\'e}line},
        title = "{Relative effect of nodes and filaments of the cosmic web on the quenching of galaxies and the orientation of their spin}",
      journal = {\aap},
         year = 2022,
        month = feb,
       volume = {658},
          eid = {A113},
        pages = {A113},
          doi = {10.1051/0004-6361/202141723},
archivePrefix = {arXiv},
       eprint = {2109.14623},
 primaryClass = {astro-ph.CO},
       adsurl = {https://ui.adsabs.harvard.edu/abs/2022A&A...658A.113M}
}

@ARTICLE{Welker2014,
       author = {{Welker}, C. and {Devriendt}, J. and {Dubois}, Y. and {Pichon}, C. and {Peirani}, S.},
        title = "{Mergers drive spin swings along the cosmic web.}",
      journal = {\mnras},
         year = 2014,
        month = nov,
       volume = {445},
        pages = {L46-L50},
          doi = {10.1093/mnrasl/slu106},
archivePrefix = {arXiv},
       eprint = {1403.2728},
 primaryClass = {astro-ph.CO},
       adsurl = {https://ui.adsabs.harvard.edu/abs/2014MNRAS.445L..46W}
}

@ARTICLE{Barsanti2025,
       author = {{Barsanti}, Stefania and {Croom}, Scott M. and {Colless}, Matthew and {Bland-Hawthorn}, Joss and {Brough}, Sarah and {Bryant}, Julia J. and {Lorente}, Nuria and {Oh}, Sree and {Santucci}, Giulia and {Sweet}, Sarah and {Sande van de}, Jesse and {Welker}, Charlotte},
        title = "{The SAMI Galaxy Survey: large-scale environment affects galaxy spin amplitudes and the formation of slow rotators}",
      journal = {\mnras},
         year = 2025,
        month = apr,
       volume = {538},
       number = {4},
        pages = {2660-2675},
          doi = {10.1093/mnras/staf426},
archivePrefix = {arXiv},
       eprint = {2503.09052},
 primaryClass = {astro-ph.GA},
       adsurl = {https://ui.adsabs.harvard.edu/abs/2025MNRAS.538.2660B}
}

@ARTICLE{Zwicky19333,
       author = {{Zwicky}, F.},
        title = "{Die Rotverschiebung von extragalaktischen Nebeln}",
      journal = {Helvetica Physica Acta},
         year = 1933,
        month = jan,
       volume = {6},
        pages = {110-127},
       adsurl = {https://ui.adsabs.harvard.edu/abs/1933AcHPh...6..110Z}
}

@ARTICLE{Weigel2018,
       author = {{Weigel}, Anna K. and {Schawinski}, Kevin and {Treister}, Ezequiel and {Trakhtenbrot}, Benny and {Sanders}, David B.},
        title = "{The fraction of AGNs in major merger galaxies and its luminosity dependence}",
      journal = {\mnras},
         year = 2018,
        month = may,
       volume = {476},
       number = {2},
        pages = {2308-2317},
          doi = {10.1093/mnras/sty383},
archivePrefix = {arXiv},
       eprint = {1802.04277},
 primaryClass = {astro-ph.GA},
       adsurl = {https://ui.adsabs.harvard.edu/abs/2018MNRAS.476.2308W}
}

@ARTICLE{Ellison2019,
       author = {{Ellison}, Sara L. and {Viswanathan}, Akshara and {Patton}, David R. and {Bottrell}, Connor and {McConnachie}, Alan W. and {Gwyn}, Stephen and {Cuillandre}, Jean-Charles},
        title = "{A definitive merger-AGN connection at z {\ensuremath{\sim}} 0 with CFIS: mergers have an excess of AGN and AGN hosts are more frequently disturbed}",
      journal = {\mnras},
         year = 2019,
        month = aug,
       volume = {487},
       number = {2},
        pages = {2491-2504},
          doi = {10.1093/mnras/stz1431},
archivePrefix = {arXiv},
       eprint = {1905.08830},
 primaryClass = {astro-ph.GA},
       adsurl = {https://ui.adsabs.harvard.edu/abs/2019MNRAS.487.2491E}
}

@ARTICLE{Darg2010,
       author = {{Darg}, D.~W. and {Kaviraj}, S. and {Lintott}, C.~J. and {Schawinski}, K. and {Sarzi}, M. and {Bamford}, S. and {Silk}, J. and {Andreescu}, D. and {Murray}, P. and {Nichol}, R.~C. and {Raddick}, M.~J. and {Slosar}, A. and {Szalay}, A.~S. and {Thomas}, D. and {Vandenberg}, J.},
        title = "{Galaxy Zoo: the properties of merging galaxies in the nearby Universe - local environments, colours, masses, star formation rates and AGN activity}",
      journal = {\mnras},
         year = 2010,
        month = jan,
       volume = {401},
       number = {3},
        pages = {1552-1563},
          doi = {10.1111/j.1365-2966.2009.15786.x},
archivePrefix = {arXiv},
       eprint = {0903.5057},
 primaryClass = {astro-ph.GA},
       adsurl = {https://ui.adsabs.harvard.edu/abs/2010MNRAS.401.1552D}
}

@ARTICLE{Alonso2007,
       author = {{Alonso}, M. Sol and {Lambas}, Diego G. and {Tissera}, Patricia and {Coldwell}, Georgina},
        title = "{Active galactic nuclei and galaxy interactions}",
      journal = {\mnras},
         year = 2007,
        month = mar,
       volume = {375},
       number = {3},
        pages = {1017-1024},
          doi = {10.1111/j.1365-2966.2007.11367.x},
archivePrefix = {arXiv},
       eprint = {astro-ph/0701192},
 primaryClass = {astro-ph},
       adsurl = {https://ui.adsabs.harvard.edu/abs/2007MNRAS.375.1017A}
}

@article{Cautun2014,
   title={Evolution of the cosmic web},
   volume={441},
   ISSN={0035-8711},
   url={http://dx.doi.org/10.1093/mnras/stu768},
   DOI={10.1093/mnras/stu768},
   number={4},
   journal={\mnras},
   publisher={Oxford University Press (OUP)},
   author={Cautun, Marius and van de Weygaert, Rien and Jones, Bernard J. T. and Frenk, Carlos S.},
   year={2014},
   month=may, pages={2923–2973} }

@ARTICLE{zoobot,
       author = {{Walmsley}, Mike and {Allen}, Campbell and {Aussel}, Ben and {Bowles}, Micah and {Gregorowicz}, Kasia and {Slijepcevic}, Inigo and {Lintott}, Chris and {Scaife}, Anna and {Jab{\l}o{\'n}ska}, Maja and {Karchev}, Kosio and {Lanzieri}, Denise and {Mohan}, Devina and {O'Ryan}, David and {Saiguhan}, Bharath and {Su{\'a}rez}, Crisel and {Guerra-Varas}, Nicol{\'a}s and {Velu}, Renuka},
        title = "{Zoobot: Adaptable Deep Learning Models for Galaxy Morphology}",
      journal = {The Journal of Open Source Software},
         year = 2023,
        month = may,
       volume = {8},
       number = {85},
          eid = {5312},
        pages = {5312},
          doi = {10.21105/joss.05312},
       adsurl = {https://ui.adsabs.harvard.edu/abs/2023JOSS....8.5312W}
}

@ARTICLE{boselli2006,
       author = {{Boselli}, Alessandro and {Gavazzi}, Giuseppe},
        title = "{Environmental Effects on Late-Type Galaxies in Nearby Clusters}",
      journal = {\pasp},
         year = 2006,
        month = apr,
       volume = {118},
       number = {842},
        pages = {517-559},
          doi = {10.1086/500691},
archivePrefix = {arXiv},
       eprint = {astro-ph/0601108},
 primaryClass = {astro-ph},
       adsurl = {https://ui.adsabs.harvard.edu/abs/2006PASP..118..517B}
}

@article{Jian2012,
   title={ENVIRONMENTAL DEPENDENCE OF THE GALAXY MERGER RATE IN A ΛCDM UNIVERSE},
   volume={754},
   ISSN={1538-4357},
   url={http://dx.doi.org/10.1088/0004-637X/754/1/26},
   DOI={10.1088/0004-637x/754/1/26},
   number={1},
   journal={\apj},
   publisher={American Astronomical Society},
   author={Jian, Hung-Yu and Lin, Lihwai and Chiueh, Tzihong},
   year={2012},
   month=jul, pages={26} }

@article{omori2023,
       author = {{Omori}, Kiyoaki Christopher and {Bottrell}, Connor and {Walmsley}, Mike and {Yesuf}, Hassen M. and {Goulding}, Andy D. and {Ding}, Xuheng and {Popping}, Gerg{\"o} and {Silverman}, John D. and {Takeuchi}, Tsutomu T. and {Toba}, Yoshiki},
        title = "{Galaxy mergers in Subaru HSC-SSP: A deep representation learning approach for identification, and the role of environment on merger incidence}",
      journal = {\aap},
         year = 2023,
        month = nov,
       volume = {679},
          eid = {A142},
        pages = {A142},
          doi = {10.1051/0004-6361/202346743},
archivePrefix = {arXiv},
       eprint = {2309.15539},
 primaryClass = {astro-ph.GA},
       adsurl = {https://ui.adsabs.harvard.edu/abs/2023A&A...679A.142O}
}

@article{Lopes2023,
       author = {{Lopes}, Paulo A.~A. and {Ribeiro}, Andr{\'e} L.~B. and {Brambila}, Douglas},
        title = "{The role of groups in galaxy evolution: compelling evidence of pre-processing out to the turnaround radius of clusters}",
      journal = {\mnras},
         year = 2024,
        month = jan,
       volume = {527},
       number = {1},
        pages = {L19-L25},
          doi = {10.1093/mnrasl/slad134},
archivePrefix = {arXiv},
       eprint = {2309.11578},
 primaryClass = {astro-ph.CO},
       adsurl = {https://ui.adsabs.harvard.edu/abs/2024MNRAS.527L..19L}
}

@article{Cox2006,
   title={The Kinematic Structure of Merger Remnants},
   volume={650},
   ISSN={1538-4357},
   url={http://dx.doi.org/10.1086/507474},
   DOI={10.1086/507474},
   number={2},
   journal={\apj},
   publisher={American Astronomical Society},
   author={Cox, T. J. and Dutta, Suvendra N. and Di Matteo, Tiziana and Hernquist, Lars and Hopkins, Philip F. and Robertson, Brant and Springel, Volker},
   year={2006},
   month=oct, pages={791–811} }

@ARTICLE{Barnes1992,
       author = {{Barnes}, Joshua E. and {Hernquist}, Lars},
        title = "{Dynamics of interacting galaxies.}",
      journal = {\araa},
         year = 1992,
        month = jan,
       volume = {30},
        pages = {705-742},
          doi = {10.1146/annurev.aa.30.090192.003421},
       adsurl = {https://ui.adsabs.harvard.edu/abs/1992ARA&A..30..705B}
}

@ARTICLE{Toomre1972,
       author = {{Toomre}, Alar and {Toomre}, Juri},
        title = "{Galactic Bridges and Tails}",
      journal = {\apj},
         year = 1972,
        month = dec,
       volume = {178},
        pages = {623-666},
          doi = {10.1086/151823},
       adsurl = {https://ui.adsabs.harvard.edu/abs/1972ApJ...178..623T}
}

@ARTICLE{Press1974,
       author = {{Press}, William H. and {Schechter}, Paul},
        title = "{Formation of Galaxies and Clusters of Galaxies by Self-Similar Gravitational Condensation}",
      journal = {\apj},
         year = 1974,
        month = feb,
       volume = {187},
        pages = {425-438},
          doi = {10.1086/152650},
       adsurl = {https://ui.adsabs.harvard.edu/abs/1974ApJ...187..425P}
}

@article{Eke1996,
    author = {Eke, Vincent R. and Cole, Shaun and Frenk, Carlos S.},
    title = {Cluster evolution as a diagnostic for Ω},
    journal = {\mnras},
    volume = {282},
    number = {1},
    pages = {263-280},
    year = {1996},
    month = {09},
    issn = {0035-8711},
    doi = {10.1093/mnras/282.1.263},
    url = {https://doi.org/10.1093/mnras/282.1.263},
    eprint = {https://academic.oup.com/mnras/article-pdf/282/1/263/18200014/282-1-263.pdf},
}

@ARTICLE{Kuchner2020,
       author = {{Kuchner}, Ulrike and {Arag{\'o}n-Salamanca}, Alfonso and {Pearce}, Frazer R. and {Gray}, Meghan E. and {Rost}, Agust{\'\i}n and {Mu}, Chunliang and {Welker}, Charlotte and {Cui}, Weiguang and {Haggar}, Roan and {Laigle}, Clotilde and {Knebe}, Alexander and {Kraljic}, Katarina and {Sarron}, Florian and {Yepes}, Gustavo},
        title = "{Mapping and characterization of cosmic filaments in galaxy cluster outskirts: strategies and forecasts for observations from simulations}",
      journal = {\mnras},
         year = 2020,
        month = jun,
       volume = {494},
       number = {4},
        pages = {5473-5491},
          doi = {10.1093/mnras/staa1083},
archivePrefix = {arXiv},
       eprint = {2004.08408},
 primaryClass = {astro-ph.GA},
       adsurl = {https://ui.adsabs.harvard.edu/abs/2020MNRAS.494.5473K}
}

@ARTICLE{Galárraga-Espinosa2020,
       author = {{Gal{\'a}rraga-Espinosa}, Daniela and {Aghanim}, Nabila and {Langer}, Mathieu and {Gouin}, C{\'e}line and {Malavasi}, Nicola},
        title = "{Populations of filaments from the distribution of galaxies in numerical simulations}",
      journal = {\aap},
         year = 2020,
        month = sep,
       volume = {641},
          eid = {A173},
        pages = {A173},
          doi = {10.1051/0004-6361/202037986},
archivePrefix = {arXiv},
       eprint = {2003.09697},
 primaryClass = {astro-ph.CO},
       adsurl = {https://ui.adsabs.harvard.edu/abs/2020A&A...641A.173G}
}

@ARTICLE{Bonjean2020,
       author = {{Bonjean}, V. and {Aghanim}, N. and {Douspis}, M. and {Malavasi}, N. and {Tanimura}, H.},
        title = "{Filament profiles from WISExSCOS galaxies as probes of the impact of environmental effects}",
      journal = {\aap},
         year = 2020,
        month = jun,
       volume = {638},
          eid = {A75},
        pages = {A75},
          doi = {10.1051/0004-6361/201937313},
archivePrefix = {arXiv},
       eprint = {1912.06559},
 primaryClass = {astro-ph.CO},
       adsurl = {https://ui.adsabs.harvard.edu/abs/2020A&A...638A..75B}
}

@ARTICLE{Haines2023,
       author = {{Haines}, C. and {Jaff{\'e}}, Y. and {Tejos}, N. and {Monachesi}, A. and {Pompei}, E. and {Finoguenov}, A. and {Sif{\'o}n}, C. and {Lopez}, S. and {Manjunatha}, A.~B. and {Bilton}, L. and {Comparat}, J. and {Cuellar}, R. and {D'Ago}, G. and {Demarco}, R. and {Lima-Dias}, C. and {L{\"o}sch}, E. and {Merluzzi}, P. and {Smith Castelli}, A. and {Sodre}, L. and {Vinicius}, E. and {CHANCES Team}},
        title = "{CHANCES: A CHileAN Cluster galaxy Evolution Survey}",
      journal = {Msngr},
         year = 2023,
        month = mar,
       volume = {190},
        pages = {31-33},
          doi = {10.18727/0722-6691/5308},
       adsurl = {https://ui.adsabs.harvard.edu/abs/2023Msngr.190...31H}
}

@article{rong2024,
       author = {{Rong}, Yu and {Shen}, Jinzhi and {Hua}, Zichen},
        title = "{Galaxy triplets alignment in large-scale filaments}",
      journal = {\mnras},
         year = 2024,
        month = jun,
       volume = {531},
       number = {1},
        pages = {L9-L13},
          doi = {10.1093/mnrasl/slae021},
archivePrefix = {arXiv},
       eprint = {2403.13273},
 primaryClass = {astro-ph.GA},
       adsurl = {https://ui.adsabs.harvard.edu/abs/2024MNRAS.531L...9R}
}

@article{Kim_2024,
   title={Distribution of Merging and Post-merger Galaxies in Nearby Galaxy Clusters},
   volume={966},
   ISSN={1538-4357},
   url={http://dx.doi.org/10.3847/1538-4357/ad32ce},
   DOI={10.3847/1538-4357/ad32ce},
   number={1},
   journal={\apj},
   publisher={American Astronomical Society},
   author={Kim, Duho and Sheen, Yun-Kyeong and Jaffé, Yara L. and Kelkar, Kshitija and Ranjan, Adarsh and Piraino-Cerda, Franco and Crossett, Jacob P. and Costa Lourenço, Ana Carolina and Martin, Garreth and Nantais, Julie B. and Demarco, Ricardo and Treister, Ezequiel and Yi, Sukyoung K.},
   year={2024},
   month=apr, pages={124} }

@article{Mesa_2018,
   title={The orientation of galaxy pairs with filamentary structures: dependence on morphology},
   volume={619},
   ISSN={1432-0746},
   url={http://dx.doi.org/10.1051/0004-6361/201832910},
   DOI={10.1051/0004-6361/201832910},
   journal={\aap},
   publisher={EDP Sciences},
   author={Mesa, Valeria and Duplancic, Fernanda and Alonso, Sol and Muñoz Jofré, Maria Rosa and Coldwell, Georgina and Lambas, Diego G.},
   year={2018},
   month=oct, pages={A24} }

@ARTICLE{Sousbie2011,
       author = {{Sousbie}, T.},
        title = "{The persistent cosmic web and its filamentary structure - I. Theory and implementation}",
      journal = {\mnras},
         year = 2011,
        month = jun,
       volume = {414},
       number = {1},
        pages = {350-383},
          doi = {10.1111/j.1365-2966.2011.18394.x},
archivePrefix = {arXiv},
       eprint = {1009.4015},
 primaryClass = {astro-ph.CO},
       adsurl = {https://ui.adsabs.harvard.edu/abs/2011MNRAS.414..350S}
}

@article{walmsley2023galaxyzoodesidetailed,
       author = {{Walmsley}, Mike and {G{\'e}ron}, Tobias and {Kruk}, Sandor and {Scaife}, Anna M.~M. and {Lintott}, Chris and {Masters}, Karen L. and {Dawson}, James M. and {Dickinson}, Hugh and {Fortson}, Lucy and {Garland}, Izzy L. and {Mantha}, Kameswara and {O'Ryan}, David and {Popp}, J{\"u}rgen and {Simmons}, Brooke and {Baeten}, Elisabeth M. and {Macmillan}, Christine},
        title = "{Galaxy Zoo DESI: Detailed morphology measurements for 8.7M galaxies in the DESI Legacy Imaging Surveys}",
      journal = {\mnras},
         year = 2023,
        month = dec,
       volume = {526},
       number = {3},
        pages = {4768-4786},
          doi = {10.1093/mnras/stad2919},
archivePrefix = {arXiv},
       eprint = {2309.11425},
 primaryClass = {astro-ph.GA},
       adsurl = {https://ui.adsabs.harvard.edu/abs/2023MNRAS.526.4768W}
}

@ARTICLE{sifón2024chanceschileanclustergalaxy,
       author = {{Sif{\'o}n}, Crist{\'o}bal and {Finoguenov}, Alexis and {Haines}, Christopher P. and {Jaff{\'e}}, Yara and {Amrutha}, B.~M. and {Demarco}, Ricardo and {Lima}, E.~V.~R. and {Lima-Dias}, Ciria and {M{\'e}ndez-Hern{\'a}ndez}, Hugo and {Merluzzi}, Paola and {Monachesi}, Antonela and {Teixeira}, Gabriel S.~M. and {Tejos}, Nicolas and {Almeida-Fernandes}, F. and {Araya-Araya}, Pablo and {Argudo-Fern{\'a}ndez}, Maria and {Baier-Soto}, Ra{\'u}l and {Bilton}, Lawrence E. and {Bom}, C.~R. and {Calder{\'o}n}, Juan Pablo and {Cassar{\`a}}, Letizia P. and {Comparat}, Johan and {Courtois}, H.~M. and {D'Ago}, Giuseppe and {Dupuy}, Alexandra and {Fritz}, Alexander and {Haack}, Rodrigo F. and {Herpich}, Fabio R. and {Ibar}, E. and {Kuchner}, Ulrike and {Lacerna}, Ivan and {Lopes}, Amanda R. and {Lopez}, Sebastian and {L{\"o}sch}, Elismar and {McGee}, Sean and {Mendes de Oliveira}, C. and {Morelli}, Lorenzo and {Moretti}, Alessia and {Pallero}, Diego and {Piraino-Cerda}, Franco and {Pompei}, Emanuela and {Rescigno}, U. and {Smith Castelli}, Anal{\'\i}a V. and {Smith}, Rory and {Sodr{\'e}}, Jr., Laerte and {Tempel}, Elmo},
        title = "{CHANCES, the Chilean Cluster Galaxy Evolution Survey: Selection and initial characterisation of clusters and superclusters}",
      journal = {\aap},
         year = 2025,
        month = may,
       volume = {697},
          eid = {A92},
        pages = {A92},
          doi = {10.1051/0004-6361/202452710},
archivePrefix = {arXiv},
       eprint = {2411.13655},
 primaryClass = {astro-ph.GA},
       adsurl = {https://ui.adsabs.harvard.edu/abs/2025A&A...697A..92S}
}

@ARTICLE{Cortese2021,
       author = {{Cortese}, L. and {Catinella}, B. and {Smith}, R.},
        title = "{The Dawes Review 9: The role of cold gas stripping on the star formation quenching of satellite galaxies}",
      journal = {\pasa},
         year = 2021,
        month = aug,
       volume = {38},
          eid = {e035},
        pages = {e035},
          doi = {10.1017/pasa.2021.18},
archivePrefix = {arXiv},
       eprint = {2104.02193},
 primaryClass = {astro-ph.GA},
       adsurl = {https://ui.adsabs.harvard.edu/abs/2021PASA...38...35C}
}

@ARTICLE{Fujita2004,
   author = {{Fujita}, Y.},
    title = "{Pre-Processing of Galaxies before Entering a Cluster}",
  journal = {\pasj},
   eprint = {arXiv:astro-ph/0311193},
     year = 2004,
    month = feb,
   volume = 56,
    pages = {29-43},
   adsurl = {http://adsabs.harvard.edu/abs/2004PASJ...56...29F}
}

@ARTICLE{Haines2015,
   author = {{Haines}, C.~P. and {Pereira}, M.~J. and {Smith}, G.~P. and 
	{Egami}, E. and {Babul}, A. and {Finoguenov}, A. and {Ziparo}, F. and 
	{McGee}, S.~L. and {Rawle}, T.~D. and {Okabe}, N. and {Moran}, S.~M.
	},
    title = "{LoCuSS: The Slow Quenching of Star Formation in Cluster Galaxies and the Need for Pre-processing}",
  journal = {\apj},
archivePrefix = "arXiv",
   eprint = {1504.05604},
     year = 2015,
    month = jun,
   volume = 806,
      eid = {101},
    pages = {101},
      doi = {10.1088/0004-637X/806/1/101},
   adsurl = {http://adsabs.harvard.edu/abs/2015ApJ...806..101H}
}

@ARTICLE{Hopkins2006,
       author = {{Hopkins}, Philip F. and {Hernquist}, Lars and {Cox}, Thomas J. and {Di Matteo}, Tiziana and {Robertson}, Brant and {Springel}, Volker},
        title = "{A Unified, Merger-driven Model of the Origin of Starbursts, Quasars, the Cosmic X-Ray Background, Supermassive Black Holes, and Galaxy Spheroids}",
      journal = {\apjs},
         year = 2006,
        month = mar,
       volume = {163},
       number = {1},
        pages = {1-49},
          doi = {10.1086/499298},
archivePrefix = {arXiv},
       eprint = {astro-ph/0506398},
 primaryClass = {astro-ph},
       adsurl = {https://ui.adsabs.harvard.edu/abs/2006ApJS..163....1H}
}

@ARTICLE{Jaffe2015,
   author = {{Jaff{\'e}}, Y.~L. and {Smith}, R. and {Candlish}, G.~N. and 
	{Poggianti}, B.~M. and {Sheen}, Y.-K. and {Verheijen}, M.~A.~W.
	},
    title = "{BUDHIES II: a phase-space view of H I gas stripping and star formation quenching in cluster galaxies}",
  journal = {\mnras},
archivePrefix = "arXiv",
   eprint = {1501.03819},
     year = 2015,
    month = apr,
   volume = 448,
    pages = {1715-1728},
      doi = {10.1093/mnras/stv100},
   adsurl = {http://adsabs.harvard.edu/abs/2015MNRAS.448.1715J}
}

@ARTICLE{White1978,
       author = {{White}, S.~D.~M. and {Rees}, M.~J.},
        title = "{Core condensation in heavy halos: a two-stage theory for galaxy formation and clustering.}",
      journal = {\mnras},
         year = 1978,
        month = may,
       volume = {183},
        pages = {341-358},
          doi = {10.1093/mnras/183.3.341},
       adsurl = {https://ui.adsabs.harvard.edu/abs/1978MNRAS.183..341W}
}

@ARTICLE{Zhou2021,
       author = {{Zhou}, Rongpu and {Newman}, Jeffrey A. and {Mao}, Yao-Yuan and {Meisner}, Aaron and {Moustakas}, John and {Myers}, Adam D. and {Prakash}, Abhishek and {Zentner}, Andrew R. and {Brooks}, David and {Duan}, Yutong and {Landriau}, Martin and {Levi}, Michael E. and {Prada}, Francisco and {Tarle}, Gregory},
        title = "{The clustering of DESI-like luminous red galaxies using photometric redshifts}",
      journal = {\mnras},
         year = 2021,
        month = mar,
       volume = {501},
       number = {3},
        pages = {3309-3331},
          doi = {10.1093/mnras/staa3764},
archivePrefix = {arXiv},
       eprint = {2001.06018},
 primaryClass = {astro-ph.CO},
       adsurl = {https://ui.adsabs.harvard.edu/abs/2021MNRAS.501.3309Z}
}

\begin{appendix}
\onecolumn
\section{Schematic view of $D_{\rm fil}$ }
\label{app:A}

\begin{figure}[htbp]
    \centering
    \includegraphics[width=0.4\textwidth]{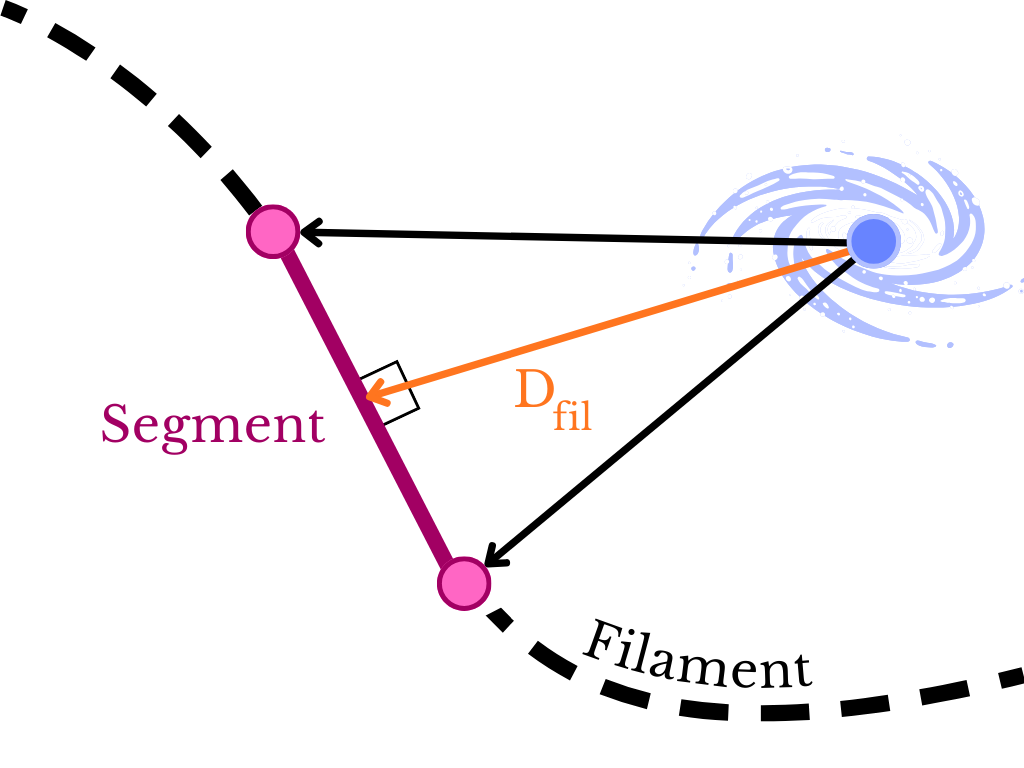}
    \caption{Schematic representation of the calculation of the minimum distance ($D_{fil}$) between a galaxy (blue) and a filament (black dashed curve). The filament is modelled as a set of linear segments (pink line between pink points). The shortest perpendicular distance is computed using the point line method, and is taken as the galaxy--filament distance.}
    \label{fig:draw}
\end{figure}

We quantified galaxy proximity to large-scale structure using the projected distance to the cluster centre ($r_{cl}$) and the shortest perpendicular distance to the nearest filament ($D_{fil}$), calculated via the point-line method where filament segments connect adjacent nodes (See Figure~\ref{fig:draw}).

In Figure~\ref{fig:draw}, we illustrate the geometric procedure used to compute the distance $D_{fil}$ between each galaxy and its nearest filament. The filaments are represented as linear segments connecting adjacent nodes, and the minimum galaxy–filament separation is derived using the point–line method described in Section 2.

\newpage
\section{Examples of merger galaxies}
\label{app:B} 

\begin{figure*}[h]
    \centering
    \includegraphics[width=0.95\textwidth]{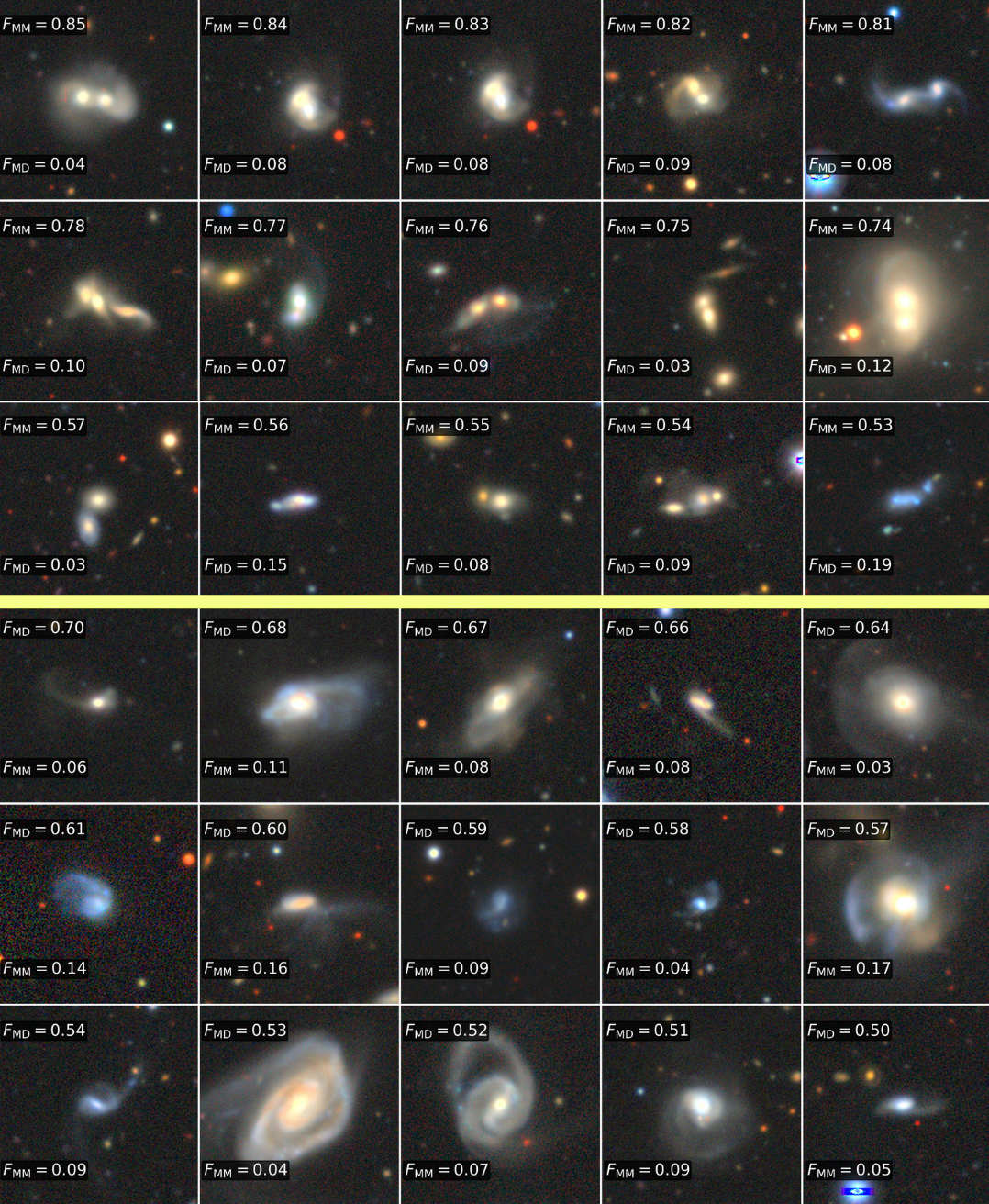}
    \caption{Examples of randomly selected merging galaxies. The upper panel shows galaxies selected by $F_{\mathrm{MM}}$, and the bottom panel shows galaxies selected by $F_{\mathrm{MD}}$. In both panels, galaxies are arranged from left to right in decreasing order of the corresponding score. 
    } 
    \label{fig:mosaic}
\end{figure*}

In Figure~\ref{fig:mosaic}, we show examples of merging galaxies identified with \textsc{Zoobot} \citep{zoobot}. The top panel displays galaxies selected according to the merger fraction ($F_{MM}$), while the bottom panel shows those selected based on the major disturbed fraction ($F_{MD}$), according to the criterion described in Section 3.

\newpage

\section{Controlling for stellar mass }
\label{app:C}

\begin{figure*}[h]
    \centering
\includegraphics[width=0.6\textwidth]{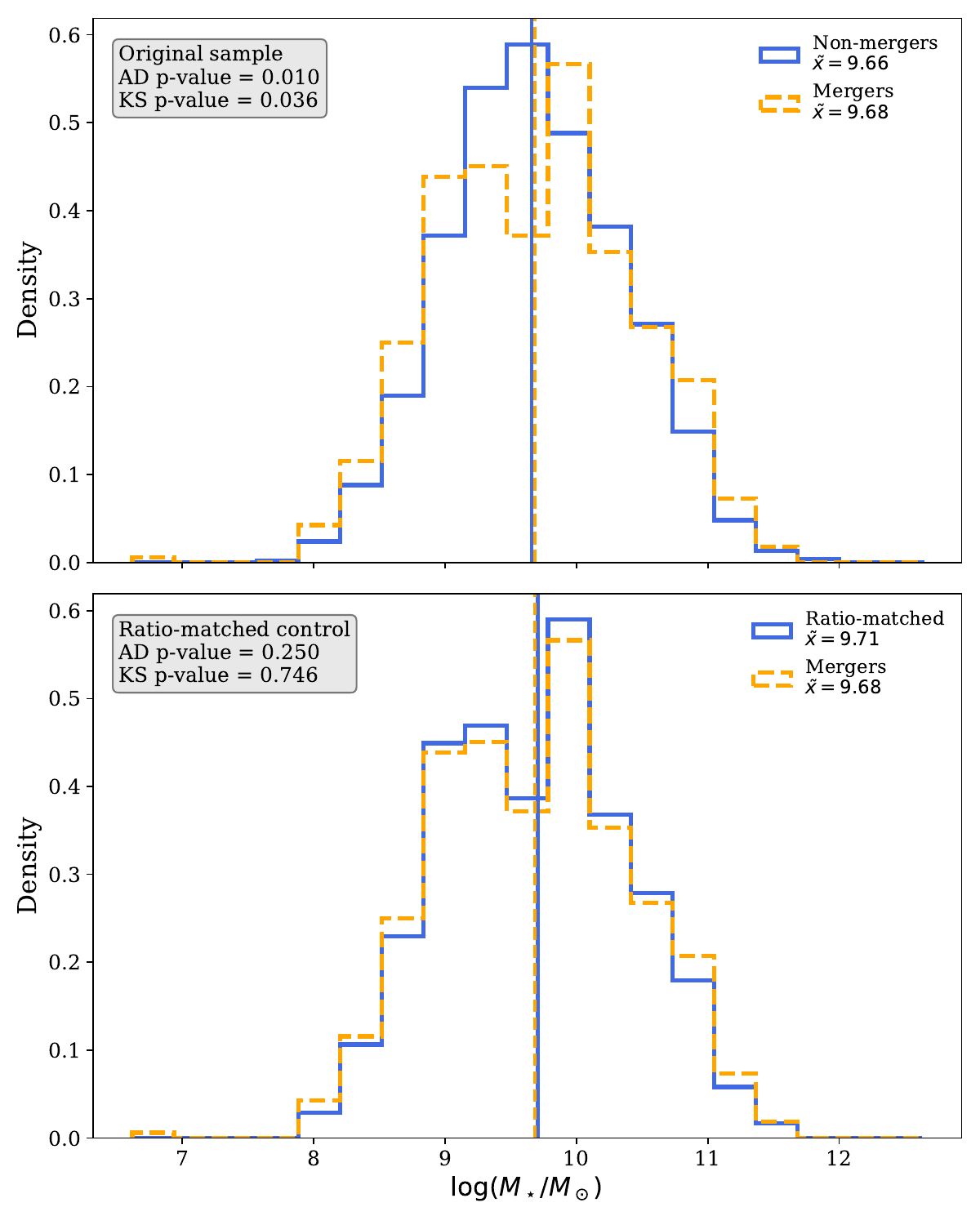}
    \caption{Stellar mass distributions of merging and non-merging galaxies. The top panel shows the original samples, which are similar in shape but statistically different (AD and KS tests). The bottom panel shows a ratio-matched control sample, where non-mergers are randomly down-sampled in stellar-mass bins to reproduce the merger mass distribution without enforcing equal sample sizes. The result corresponds to one realization from 1000 Monte Carlo resamplings. Controlling for stellar mass does not change the conclusions.} 
    \label{fig:mass_matched}
\end{figure*}

To check that our results are not driven by stellar mass, we have compared the stellar mass distribution of the merger and non-merger galaxy samples.
Stellar masses ($M_{\star}$) were computed following the empirical relation between rest-frame ($g - i$) colour and M$_\star$/L$_i$ in \citet{Taylor2011} for galaxies with reliable photometric measurements (76\% of the sample). The remaining objects lack complete photometric coverage in the required filters and therefore lack a stellar-mass estimate.
The top panel of Figure~\ref{fig:mass_matched} shows that, while the 2 distributions look similar, statistically they are likely to be drawn from the same parent distribution. 
We checked how these differences could affect our results by repeating the analysis using mass-matched samples. To control for stellar mass without significantly reducing the non-merger sample size, we construct a ratio-matched control sample. Specifically, we bin both populations in stellar mass and randomly down-sample the non-merging galaxies within each mass bin according to the relative merger-to-non-merger ratio, so as to reproduce the shape of the merger mass distribution without enforcing equal sample sizes or up-sampling (see bottom panel of Figure~\ref{fig:mass_matched}). This procedure is repeated 1000 times using Monte Carlo. Using this mass-controlled sample, we find that the statistical difference in filament distances between mergers and non-mergers remains unchanged.

\newpage
\section{Alignment of mergers with filaments in the entire cluster sample}
\label{app:D} 

\begin{figure*}[h]
    \centering
\includegraphics[width=0.9\textwidth]{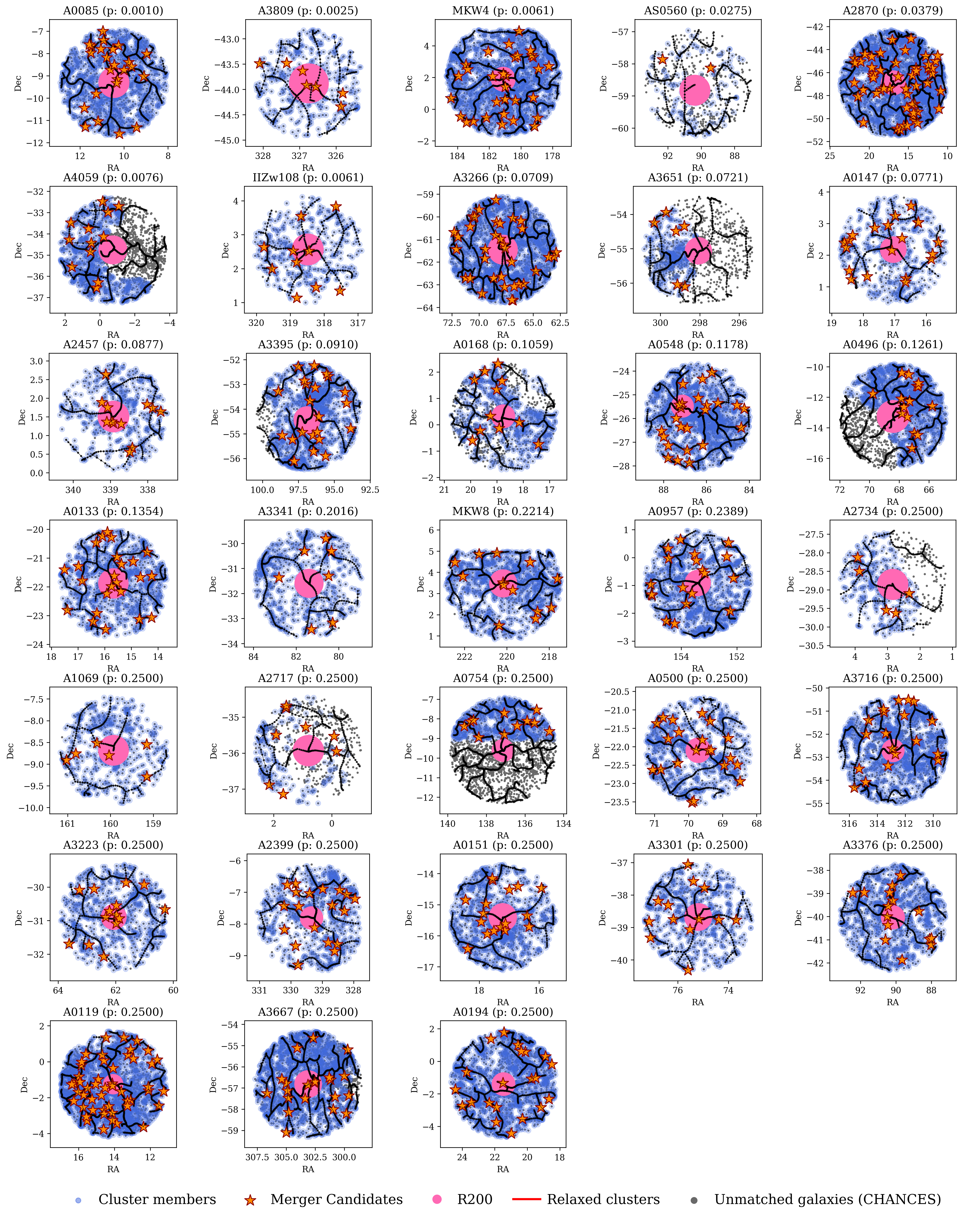}
    \caption{Distribution of galaxy-galaxy mergers (orange stars) across the 33 clusters in our sample. Black lines represent filaments, the pink circle denotes the virial radius ($R_{200}$), and blue points correspond to the general cluster galaxy population. For visualization purposes, grey points represent all unmatched cluster members from the CHANCES catalogue, which were not included in the statistical analysis.  }
    \label{fig:all_mosaic}
\end{figure*}

\begin{figure*}[h]
    \centering
\includegraphics[width=1\textwidth]{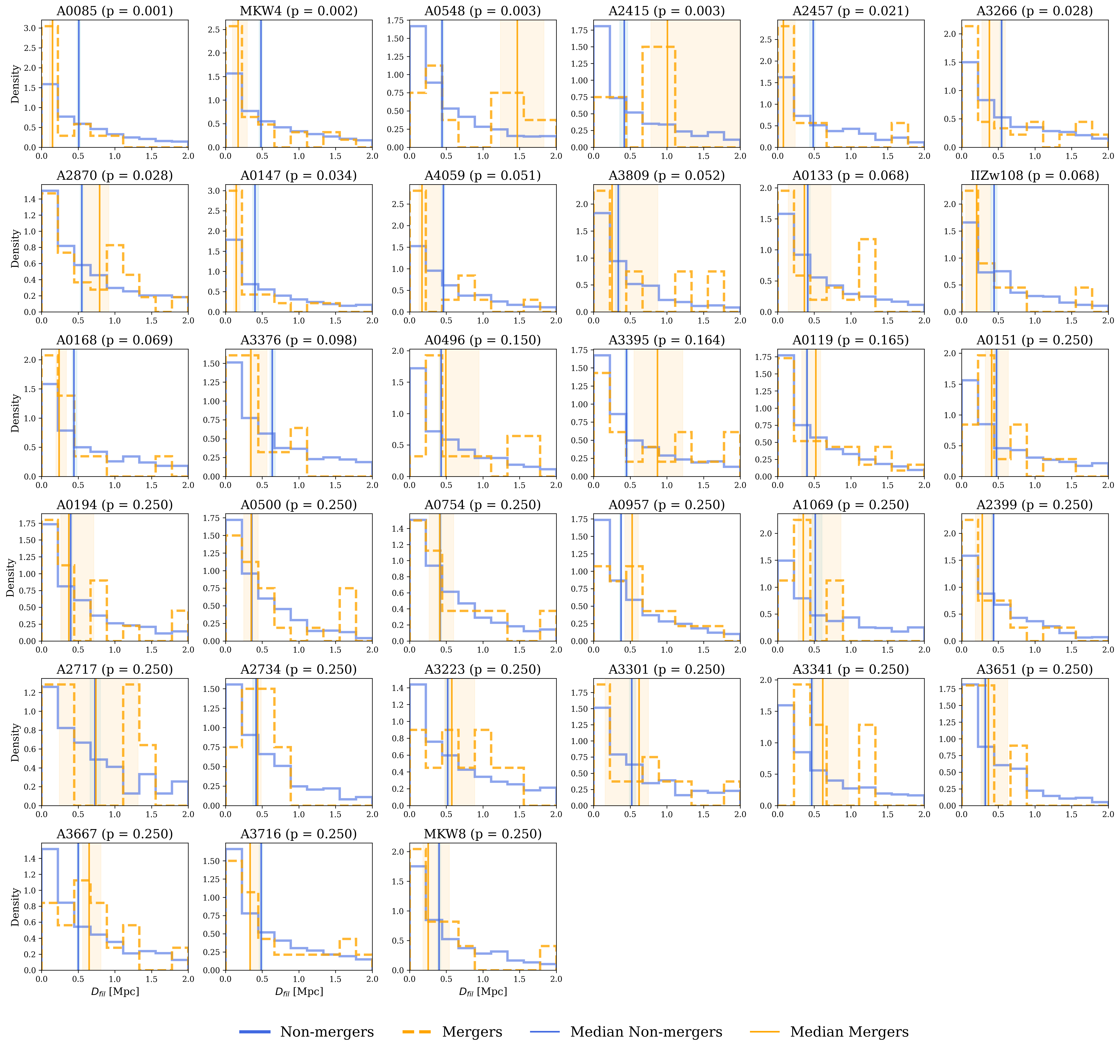}
    \caption{Normalized distribution of the nearest distances to filaments for galaxy mergers located outside $1 \times R_{200}$ in the 33 galaxy clusters (orange dashed line), compared to the non-merging cluster galaxy population outside $1 \times R_{200}$ of each respective cluster (blue dotted line). Error bars correspond to Poisson uncertainties in each bin, while vertical lines mark the median of each distribution. Shaded regions indicate the 68\% bootstrap confidence intervals (percentiles 16–84) around the median. Clusters are ordered from lowest to highest p-value, where a lower p-value indicates stronger evidence of a difference between the distributions. Distances to filaments are truncated at 2 Mpc to emphasize the vicinity where most merger candidates are found, although the full distributions extend up to $\sim$4 Mpc.} 
    
    \label{fig:outside}
\end{figure*}

Figures~\ref{fig:all_mosaic} and~\ref{fig:outside} present the clusters used in our analysis, ordered by increasing $p$-value. 
The first figure shows the spatial distribution of merger candidates along the filaments. 
The second figure displays the corresponding distributions of distances to the filament for each cluster.

\end{appendix}
\twocolumn
\end{document}